\documentstyle[aps,preprint,tighten]{revtex}

\begin{document}

\draft

\title{
Dynamic Renormalization Group Approach to Self-Organized
Critical Phenomena
}

\author{Albert D\'{\i}az-Guilera}
\address{
Departament de F\'{\i}sica Fonamental \\ Facultat de F\'{\i}sica,
Universitat de Barcelona \\ Diagonal 647, E-08028 Barcelona, Spain }

\maketitle

\begin{abstract}

Two different models exhibiting self-organized criticality are
analyzed by means of the dynamic renormalization group.  Although the
two models differ by their behavior under a parity transformation of
the order parameter, it is shown that they both belong to the same
universality class, in agreement with computer simulations. The
asymptotic values of the critical exponents are estimated up to one
loop order from a systematic expansion of a nonlinear equation in the
number of coupling constants.

\end{abstract}

\narrowtext

        Recently much attention has been paid to the phenomenon known
as self-organized criticality. Bak, Tang, and Wiesenfeld
\cite{pra38.364} studied a cellular automaton model as a paradigm for
the explanation of two widely occurring phenomena in nature: $1/f$
noise and fractal structures.  Both have in common a lack of
characteristic scales.  This scale invariance suggests that the
system is critical in analogy with classical equilibrium critical
phenomena, but in self-organized criticality one deals with dynamical
nonequilibrium statistical properties.  On the other hand the system
evolves naturally to the critical state without any tuning of
external parameters.

        Several cellular automata models exhibiting self-organized
criticality have been reported in the literature. In the original
sandpile model of Bak {\em et al.} \cite{pra38.364} the system is
perturbed externally by a random addition of sand grains. Once the
slope between two contiguous cells have reached a threshold
condition, sand is transferred to its neighbors in a fixed amount.
Taking this model as a reference, different dynamical rules have been
investigated leading to a wide variety of universality classes.
Some authors have attempted to connect
these models to stochastic
differential equations
\cite{prl62.1813,prl64.1927}. These continuous
descriptions are built according to the symmetry rules obeyed by the
discrete models. Nevertheless none of them explicitly includes the
threshold condition. On the other hand anomalous diffusion equations with
singularities in the diffusion coefficient have been considered in
order to study the deterministic dynamics of the avalanches generated
in the critical state
\cite{prl65.2547,prl68.2058}.

        In this paper we study two nonlinear stochastic differential
equations derived from the discrete dynamical rules of two models
which have different symmetry properties
from which one expects different critical behavior, but in the
simulations they give the same critical exponents. We will show
analytically, by means of the dynamic renormalization group (DRG)
\cite{pra16.732,pra39.3053}, that both models belong to
the same universality class. The threshold condition is kept but the
step function is regularized in order to allow a power series
expansion. It is in the limit where the threshold is recovered that
the coupling constants that distinguish both models are shown to be
irrelevant. Once this equivalence is established we deal with one of
the models, showing that an infinite number of coupling constants is
relevant below the upper critical dimension $d_c=4$; by expanding in
the number of nonlinearities to first order in $\varepsilon=4-d$, we
get an accurate estimate of the dynamical exponent close to the value
obtained in the numerical simulations.

        First, we describe briefly the dynamics of the two models
showing self-organized criticality. The first model, to be referred
to as model A, was originally proposed by Zhang \cite{prl63.470} and
it consists of a lattice in which any site can store some energy $E$
continuously distributed between $0$ and $E_c$. This variable which
we call energy can have different physical interpretations
\cite{p173a.22}. The system is perturbed by adding a random amount of
energy $ \delta E$ at a randomly chosen site. Once a site reaches a
value of the energy greater than some critical value $E_c$ this site
becomes active and transfers the energy
to its nearest neighbors. At this point the input of energy from the
outside is turned off and the energy transferred to the neighboring
sites can make them active
giving rise to an activation cluster or avalanche that ends when all
the sites have reached a value of the energy smaller than $E_c$. It
is only when the avalanche has stopped that energy is added again,
otherwise the system remains quiescent.
Model B is closer
to the original sandpile model of Bak {\em et al.}
\cite{pra38.364}
and differs from A in that the amount of energy
a critical site transfers to its neighbors is a fixed amount $E_c$
instead of its whole energy $E$.

The microscopic rules for these models can be written in the form of
a set of algebraic equations and coarse-grained to get effective
equations in terms of a rescaled energy $E-E_c\rightarrow E$
as follows
\begin{mathletters}
\label{all-cont}
\begin{equation}
\frac{ \partial E({\bf \mbox{r}},t)}{ \partial t}=
\alpha  \nabla ^{2}\left[  \theta (E({\bf \mbox{r}},t))
\left( E({\bf \mbox{r}},t)+E_c
\right) \right]
+ \eta ({\bf \mbox{r}},t),
\label{z-cont}
\end{equation}
and
\begin{equation}
\frac{ \partial E({\bf \mbox{r}},t)}{ \partial t}=
\alpha  \nabla ^{2}\left[  \theta (E({\bf \mbox{r}},t)) E_c \right]
+ \eta ({\bf \mbox{r}},t)
\label{btw-cont}
\end{equation}
\end{mathletters}
for models A and B, respectively. In these equations $ \alpha $
plays the role of a diffusion constant. The threshold condition
enters through the Heaviside step function $ \theta $.
The external noise has a zero mean value and a correlation
function given by
\begin{equation}
< \eta ({\bf \mbox{r}},t) \eta ({\bf \mbox{r}}',t')> = 2 \Gamma
\delta ^d({\bf \mbox{r}}-{\bf \mbox{r}}')
\label{external}
\end{equation}
as was extensively discussed in \cite{pra45.8551}.

In order to perform a perturbative expansion of Eq.  (\ref{all-cont})
we choose one of the representations of the step function
\cite{prl68.2058}
\begin{equation}
  \theta (x) = \lim_{ \beta  \rightarrow  \infty } \frac{1}{ \pi }
\left[ \frac{ \pi }{2}+\arctan  \beta x
\right]
\label{theta}
\end{equation}
and make a series expansion of the $\arctan$ in powers of its
argument.

We can then formally write for both models
\begin{equation}
 \frac{ \partial E({\bf \mbox{r}},t)}{ \partial t}=D \nabla
^{2}E({\bf \mbox{r}},t)+ \sum_{n=2}^ \infty   \lambda _n  \nabla
^{2}E^n({\bf \mbox{r}},t)+ \eta ({\bf \mbox{r}},t) \label{series}
\end{equation}
where the effective diffusion constant ($D$) and the coupling
constants ($ \lambda _n$) take different values depending on the
symmetry of the model
\begin{equation}
D^A= \alpha \left( \frac{ \beta E_c}{ \pi }+ \frac{1}{2}  \right) ;
D^B=\frac{ \alpha  \beta E_c}{ \pi }
\end{equation}
\begin{equation}
 \lambda ^A_{2n}=-\frac{(-1)^n  \alpha  E_c  \beta ^{2n-1}}{ \pi
(2n-1)};\:\:\: \lambda ^B_{2n}=0;\:\:\:
 \lambda ^A_{2n-1}= \lambda ^B_{2n-1}=-\frac{(-1)^n  \alpha  E_c
\beta ^{2n-1}}{ \pi (2n-1)}
\end{equation}

The difference between both models is that
B has a definite parity under $E \rightarrow -E$ because for this
model the even coupling constants vanish.  Moreover even and odd
coupling constants depend on $ \beta $ in a different way. This fact
will be shown to be crucial as far as universality classes are
concerned since for the original models $ \beta  \rightarrow  \infty
$. In this limit we can see that $D^A=D^B=D$.

        The scaling behavior of
(\ref{all-cont}) is obtained from
the correlation function
\begin{equation}
<\left( E({\bf \mbox{r}}_0,t_0) - E({\bf \mbox{r}}_0+{\bf
\mbox{r}},t_0+t) \right)^{2} >^{ \frac{1}{2} } \sim  r^{ \chi }
f(t/r^z)
\label{scaling}
\end{equation}
where $ \chi $ and $z$ are the roughening and
dynamical exponents, respectively. The relevance of the different
coupling constants can be checked by naive dimensional analysis: a
change of scale ${\bf \mbox{r}}\rightarrow e^l {\bf \mbox{r}}$ is
followed by $t \rightarrow e^{zl}t$ and $E \rightarrow e^{ \chi l}E$
in order for (\ref{scaling}) be satisfied. Under this scaling
transformation $z$ and $ \chi $ are chosen to keep the linear
equation invariant. With these values one can check that all
dimensionless coupling constants are of the same order in $
\varepsilon =4-d$ and hence all nonlinear terms are equally relevant
for $d<4$.

The dynamic renormalization group procedure consists of two steps: an
elimination of the fast short wave-length modes followed by a
rescaling of the remaining modes
\cite{pra16.732,pra39.3053}. The recursion
relations after an infinitesimal transformation for the coefficients
are, to one loop order \cite{p.albert},
\begin{mathletters}
\begin{equation}
\frac{dD}{dl}=D \left[ z-2+3\frac{ \Gamma  \Lambda ^{d-4}
\lambda _3}{D^3}-
4\frac{ \Gamma  \Lambda ^{d-4} \lambda _2^2}{D^4} \right]
\end{equation}
\begin{equation}
\frac{d \Gamma }{dl}= \Gamma  \left[ 2z-2 \chi -d \right]
\label{dgdl}
\end{equation}
\begin{equation}
\frac{d \lambda _n(l)}{dl}= \lambda _n\left[ (n-1) \chi +z-2)
\right]
+ \sum_{j}\gamma_j  \Gamma  \Lambda ^{d-4}D^{a_j}\prod_{i=2}^{n+2}
\lambda _i^{b_{ij}}
\label{el}
\end{equation}
\end{mathletters}
where $\gamma_j$ are numerical factors and $b_{ij}$ are natural
numbers accounting for the number of times the bare vertex $ \lambda
_i$ enters the renormalization of the vertex $ \lambda _n$ in the
$j$-th term of the sum.  By simple dimensional analysis we get
$a_j=-1-\sum b_{ij}$ and $\sum (i-1)b_{ij}=n+1$ obtaining that the
terms within the sum of (\ref{el}) behave as $ \beta
^{n-\sum_{\text{e}}b_{ij}}$ where the sum runs over even values of
$i$. Thus we obtain
\begin{equation}
\frac{1}{ \lambda _n}\frac{d \lambda _n}{dl}  \sim
\left\{
\begin{array}{ll}
 \beta ^{1-\sum_{\text{e}}b_{ij}} & \;\; \mbox{n even} \\ \beta
^{-\sum_{\text{e}}b_{ij}} & \;\; \mbox{n odd}
\end{array}
\right.
\end{equation}
And from this behavior we can conclude: a) for odd $n$ any term
containing even vertices will not contribute when compared with terms
involving only odd vertices when $\beta \rightarrow \infty$, in
particular $ \lambda _3$ and $D\equiv \lambda _1$ get decoupled from
all even coupling constants; b) for even $n$ only those terms with
one even vertex will contribute
\footnote{The parity of a renormalized vertex is $-(-1)^e$ where
$e$ is the
number of even bare vertices. Thus an even renormalized vertex always
contains an odd number of even bare vertices.}.
In this case we can ensure that there are no
divergences caused by the $\beta \rightarrow \infty$ limit in the
equations for the even coupling constants and thus no even terms are
generated under renormalization in the equations for the odd coupling
constants.  The critical exponents are calculated at the fixed points
and from the previous arguments they will only depend on $ \lambda
_3$ which in turn is only coupled to odd coupling constants. We can
conclude finally that the values of the critical exponents will be
the same irrespective of the values taken by the even nonlinearities
showing thus the equivalence between the Zhang and BTW models.
After having shown that both models belong to the same universality
class
we will focus our analysis on simpler model B.

Our starting point is Eq. (\ref{series}) with $ \lambda _{2n}=0$.
Since all terms in this expansion are equally relevant below the
upper critical dimension ($d_c=4$) a calculation of the critical
exponents would involve all the contributions. Nevertheless the
exponents can be obtained to first order in $\varepsilon$ as a series
in the number of coupling constants which are taken into account.

One starts with the simpler case
\begin{equation}
 \frac{ \partial E({\bf \mbox{r}},t)}{ \partial t}=D \nabla
^{2}E({\bf \mbox{r}},t)+ \lambda _3  \nabla ^{2}E^3({\bf
\mbox{r}},t)+ \eta ({\bf \mbox{r}},t)
\label{l3}
\end{equation}
where $D=\frac{ \alpha  \beta E_c}{ \pi }$ and $ \lambda _3=-\frac{
\alpha E_c \beta ^3}{3 \pi }$.

By application of the DRG procedure we get the following differential
equation for the effective dimensionless coupling constant $\bar{
\lambda }_3=\frac{ \Gamma  \Lambda ^{d-4} \lambda _3}{D^3}$
\cite{pra45.8551}
\begin{equation}
\frac{d\bar{ \lambda }_3}{dl}=\bar{ \lambda }_3 \left[ \varepsilon-27
\bar{ \lambda }_3 \right]
\label{l3e}
\end{equation}
subjected to the initial condition $\bar{ \lambda }_3^0=-\frac{ \pi
^{2} \Gamma  \Lambda ^{d-4}}{3 \alpha ^{2}E_c^{2}}$. From (\ref{l3e})
we get the following behavior in $\bar{ \lambda }_3$ space.  For
$d<4$ one can expect different behavior depending on the initial sign
of $\bar{ \lambda }_3$. When it is positive it flows towards a stable
fixed point $\bar{ \lambda }_3^*=\varepsilon /27$ giving a dynamical
exponent $z=2-\varepsilon /9$. A negative $\bar{ \lambda }_3$ flows
away giving rise in principle to a strong coupling fixed point.  The
exact solution of Eq.  (\ref{l3e}) shows that $\bar{ \lambda }_3$,
even for $\bar{ \lambda }_3^0<0$, flows back to the stable fixed
point. But in this case the physical relevance of the stable fixed
point is a conjecture since along the flow in parameter space the
effective coupling constant should stay of order $ \varepsilon $. On
the other hand, above the upper critical dimension the system evolves
towards the trivial fixed point $\bar{ \lambda }_3=0$ giving a
diffusive behavior with $z=2$.

The next step is to study the contribution of $\lambda_5$ to the
critical exponents.
The equation to study is equivalent to (\ref{l3}) up to
fifth order.
In this case the differential recursion
relations for the effective coupling constants
\begin{equation}
\bar{ \lambda }_3=\frac{ \Gamma  \Lambda ^{d-4} \lambda _3}{D^3};
\hspace{1em}
\bar{ \lambda }_5=\frac{ \Gamma  \Lambda ^{d-4} \lambda _5}{D^2
\lambda _3}
\end{equation}
are \cite{p.albert}
\begin{equation}
\frac{d\bar{ \lambda }_3}{dl}=\bar{ \lambda }_3 \left[ \varepsilon-27
\bar{ \lambda }_3 +10\bar{ \lambda }_5\right]
\hspace{2ex} \mbox{and} \hspace{2ex}
\frac{d\bar{ \lambda }_5}{dl}=\bar{ \lambda }_5 \left[ \varepsilon-78
\bar{ \lambda }_3 -10\bar{ \lambda }_5 +81 \frac{\bar{ \lambda
}_3^2}{\bar{ \lambda }_5}\right]
\end{equation}
with initial conditions
$
\bar{ \lambda }_3^0=-\frac{ \pi ^{2} \Gamma  \Lambda ^{d-4}}{3 \alpha
^{2}E_c^{2}}
$
and
$
\bar{ \lambda }_5^0=-\frac{3 \pi ^{2} \Gamma  \Lambda ^{d-4}}{5
\alpha ^{2}E_c^{2}}=\frac{9}{5}\bar{ \lambda }_3^0.
$.

For $d<4$ and $\bar{ \lambda }_3^0 >0$ the solution of these
equations flows towards a new stable fixed point whose projection
onto the $\bar{ \lambda }_3$ axis is shifted with respect to the one
calculated only with the third-order nonlinearity, giving a different
dynamical exponent
\begin{equation}
z=2-\frac{159+3\sqrt{1009}}{1350}\varepsilon.
\label{z35}
\end{equation}
Nevertheless for our model with negative initial values of $\bar{
\lambda }_3$ and $\bar{ \lambda }_5$ the effective coupling constants
flow away to a strong coupling fixed point under RG transformations.
The same line of reasoning as for $\bar{ \lambda }_3$ can be applied
to the present case and the values of the critical exponents can be
obtained from the stable fixed point.

We have performed calculations with $ \lambda _7$ and $ \lambda _9$
which give further corrections to the dynamical exponent
\cite{p.albert}.
The results are
presented in Fig. 1.  where one can see that the asymptotic behavior
approaches the values obtained in \cite{prl63.470} by means of
scaling arguments and to the results from computer simulations
\cite{pra45.8551}.  The agreement is much better
in $d=3$ than in $d=2$ since we are performing a one-loop expansion.
Up
to the above mentioned order we have found that all possible choices
of the dimensionless coupling constants have an initial value which
is independent of $\beta$, legitimizing thus the expansion of the
step function (\ref{theta}).

To summarize, we have studied the critical exponents of two
models which show self-organized criticality. From the microscopic
rules one writes effective long wave-length equations involving the
threshold condition. After a power series expansion of the threshold
condition, these equations are suitable for an application of the
DRG. This allows us to show that both models are in the same
universality class and to get an estimate for the dynamical exponents
to order $\varepsilon$ by expanding in the number of coupling
constants.
The application of this technique should also be useful for
other kinds of problems in which an infinite number of relevant
nonlinearities arises.

The author wishes to acknowledge J. Bonet Avalos for very fruitful
discussions and A.-M.S. Tremblay for a critical reading of the
manuscript. This work has been supported by CICyT of the Spanish
Government, grants \#PB89-0233 and \#PB92-0863.

\begin{figure*}
\caption{Dynamical exponent as a function of $1/n$, where $n$ is the
highest odd vertex used to compute the dynamical exponent to one-loop
order.  Circles and squares are the analytical results, dashed lines
are best linear fits, and the crosses are the results from [8],
$z=\frac{d+2}{3}$.}
\end{figure*}

\end{document}